# NRBdMF: A recommendation algorithm for predicting drug effects considering directionality

Iori Azuma[1, *]; Tadahaya Mizuno[1, *, †]; Hiroyuki Kusuhara[1]

[1] Department of Pharmaceutical Sciences, The University of Tokyo, Bunkyo, Tokyo, Japan

† Corresponding author: Tadahaya Mizuno, tadahaya@gmail.com

Graduate School of Pharmaceutical Sciences, the University of Tokyo, Bunkyo-ku, Tokyo, 113-0033, Japan

Tel: +81-3-5841-4771

*These authors contributed equally.

**Declaration of competing interest**

The authors declare that they have no known competing financial interests or personal relationships that could have appeared to influence the work reported in this paper.

**Funding**

This work was supported by the JSPS KAKENHI Grant-in-Aid for Scientific Research (C) (grant number 21K06663) from the Japan Society for the Promotion of Science.

ABSTRACT

Predicting the novel effects of drugs based on information about approved drugs can be regarded as a recommendation system. Matrix factorization is one of the most used recommendation systems and various algorithms have been devised for it. A literature survey and summary of existing algorithms for predicting drug effects demonstrated that most such methods, including neighborhood regularized logistic matrix factorization, which was the best performer in benchmark tests, used a binary matrix that considers only the presence or absence of interactions. However, drug effects are known to have two opposite aspects, such as side effects and therapeutic effects. In the present study, we proposed using neighborhood regularized bidirectional matrix factorization (NRBdMF) to predict drug effects by incorporating bidirectionality, which is a characteristic property of drug effects. We used this proposed method for predicting side effects using a matrix that considered the bidirectionality of drug effects, in which known side effects were assigned a positive (+1) label and known treatment effects were assigned a negative (−1) label. The NRBdMF model, which utilizes drug bidirectional information, achieved enrichment of side effects at the top and indications at the bottom of the prediction list. This first attempt to consider the bidirectional nature of drug effects using NRBdMF showed that it reduced false positives and produced a highly interpretable output.

*Keywords*:

Matrix factorization

Recommendation system

Side effects

Therapeutic indications

**1. Introduction**

Prediction of drug effects, including target protein effects, side effects, and drug–drug interactions, is important in drug discovery, which is a primary objective of pharmaceutical sciences. Considerable information is available regarding the safety, efficacy, and tolerance of approved drugs, which can minimize the expenditure and time required to predict new drug effects. Therefore, methods for predicting unrecognized drug effects using the information on existing drugs are now attracting attention.

Drug effect prediction can be regarded as a recommendation task that suggests candidates for potential effects. With the development of machine learning techniques, several studies have investigated recommendation systems in recent years. Moreover, the application of recommendation systems based on matrix factorization for the prediction of drug effects has been well studied and various algorithms have been devised[1–10]. However, these methodologies are currently not organized and integrated knowledge about the difference in properties and concept of each method is not yet available.

In the present study, we first conducted a survey of existing drug effect prediction algorithms using matrix factorization recommendation systems, focusing on their conceptual aspects. We found that existing methods can be broadly classified into seven representative algorithms. We next compared the prediction performance of these methods. Among the representative methods, neighborhood regularized logistic matrix factorization (NRLMF) [7] showed the best performance. However, most existing methods, including these representative algorithms, use a matrix with the value "1" for observed interactions and "0" for others. This indicates that only the presence or absence of interaction is considered. Moreover, drug effects are known to have two opposite aspects, such as agonism/antagonism and therapeutic effects/side effects [11,12]. In addition, no recommendation method that can handle directionality, such as the bidirectional nature of drug effects, currently exists. Incorporation of directionality into a recommendation algorithm is expected to improve the representation capability of the model, which would increase the interpretability of results and reduce false positives in predicting drug effects due to the bilateral character (i.e., a therapeutic effect one time and a side effect the other time) [12].

Based on the survey findings, we proposed a neighborhood regularized bidirectional matrix factorization (NRBdMF) method to predict drug effects while considering positive and negative bidirectional effects. The proposed algorithm handles categorical data with directionality rather than binary data of 0 and 1, which are a kind of ordinal scale data but slightly different in that the algorithm assumes symmetric data centered at 0. In addition, we applied this proposed method to predicting side effects using a matrix that considered the duality of drug effects, in which known side effects were assigned a positive (+1) label and known treatment effects were assigned a negative (−1) label. The performance of a novel side effect prediction using NRBdMF was compared with that using NRLMF, which is one of the representative algorithms that performed well throughout the survey. Because of the ability of NRBdMF to consider bidirectionality, fewer false positives were obtained in the prediction of side effects than those obtained using NRLMF.

The present study has two contributions to the field of drug effect prediction. We surveyed studies using matrix factorization to predict drug effects and compared the performance of seven representative methods. Inspired by NRLMF, which showed excellent performance, we proposed NRBdMF and applied it to predict side effects by considering the side effects and indications as positive and negative directions of drug effects. This first attempt using the NRBdMF method showed that it reduced false positives and produced a highly interpretable output.

## 2. Materials and methods

### 2.1. Survey of existing algorithms

We followed the PRISMA guidelines and investigated studies that used matrix factorization to predict the relationship between drugs and their effects [13]. The selection of studies was based on the principles presented in Table S1. We classified the selected studies and found seven representative algorithms that were rich in derivations: NMF, GRNMF, LMF, NRLMF, CMF, TMF, and IMC [4–10]. We implemented these algorithms in Python and made them comparable by running them in the same environment. All relevant codes will be made available for public access at https://github.com/mizuno-group/NRBdMF at the time of publication.

*2.2. Comparison of representative algorithms*

We compared the prediction performance of the seven representative algorithms using two benchmark datasets, including protein and disease as drug targets (see Table 1 for details). The performances were evaluated using 10-fold cross-validation (CV) performed five times under three different settings: CVS1, CVS2, and CVS3. In CVS1, drug–target pairs in the test set were randomly selected for prediction as the standard evaluation setting. In CVS2 and CVS3, random rows or columns were blinded for testing, assuming to evaluate the ability to predict interactions with novel drugs and novel targets. The entire drug profile is selected as the test set in CVS2 and the entire target profile in CVS3. Because CVS2 and CVS3 require prior information as input, they were applied only for the three algorithms (i.e., NRLMF, TMF, and IMC) in the present study. The CVS1, CVS2, and CVS3 were settings for the prediction of new pairs, drugs, and targets, respectively. These three performance evaluation scenarios were found to be used widely [14–17]. This evaluation metric was the most used throughout the survey.

The parameter of tolerance was optimized by selecting from $\{10^{-4}, 10^{-3}, 10^{-2}, 10^{-1}, 10^{0}\}$, while default values were used for the other parameters.

**Table 1.** Summary of the benchmark datasets.

|  | Drug-Protein | Drug-Disease |
| --- | --- | --- |
| Number of drugs | 708 | 663 |
| Number of targets | 1512 | 409 |
| Number of interaction pairs | 1332 | 2532 |
| Sparsity of the interaction matrix | 0.124% | 0.934% |

## 2.3. Benchmark datasets

### 2.3.1. Drug–protein matrix

The associations between drugs and target proteins were downloaded from Luo's work [18]. A total of 1332 drug–protein associations between 708 drugs and 1512 proteins were obtained.

### 2.3.2. Drug–disease matrix

A database for associations between drugs and diseases called Cdataset was downloaded from Luo's work [19]. A total of 2532 drug–disease associations between 663 drugs and 409 diseases were obtained.

## 2.4. Kernel preparation

NRLMF and CMF require similarity square kernels for execution. We downloaded each drug and protein similarity kernel from Luo's work [18]. The disease similarity kernels were downloaded from Liang work [20]. Furthermore, TMF and IMC need low-dimensional vector representation as kernel information. Following Luo's method, the desired vectors were obtained by performing a random walk with restart method and diffusion component analysis [18]. A total of 100, 400, and 50 dimensions were selected for drugs, proteins, and diseases, respectively, in both benchmark datasets as the number of dimensions for the low-dimensional vectors. The determination of the number of dimensions was performed in accordance with Luo's work [18].

## 2.5. NRBdMF

Inspired by NRLMF [7], we developed a multilabel learning algorithm named NRBdMF that can handle directionality and predict potential drug–target interaction.

In the present study, the observed label matrix is denoted by $\boldsymbol{Y} \in \mathbb{R}^{m \times n}$, where $m$ and $n$ are the number of drugs and targets, respectively. Each element is denoted by $y_{ij} \in \{-1, 0, +1\}$, where $+1$ denotes positive labels, $-1$ denotes negative labels, and $0$ denotes missing labels. We decomposed the interaction matrix $\boldsymbol{Y}$ into two low-rank latent feature matrices $\boldsymbol{U} \in \mathbb{R}^{m \times k}$ and $\boldsymbol{V} \in \mathbb{R}^{n \times k}$, where $k$ is the dimension of the latent feature vectors. We assumed that $\boldsymbol{Y}$ can be represented by the product of $\boldsymbol{U}$ and $\boldsymbol{V}$ as follows:

$$\arg\min_{\boldsymbol{U},\boldsymbol{V}} \|\boldsymbol{R} \circ (\boldsymbol{Y} - \boldsymbol{U}\boldsymbol{V}^T)\|_F^2 \qquad (1)$$

where $\|\cdot\|_F$ denotes the Frobenius norm and ° denotes the Hadamard product of two matrices. Let $\boldsymbol{R} \in \mathbb{R}^{m \times n}$ be the indicator matrix where $r_{ij} = pw$ when $y_{ij} = 1$, $r_{ij} = nw$ when $y_{ij} = -1$, and 0 otherwise. Note that *pw* and *nw* were the weights of the positive and negative labels, respectively, and the default values were both 1. With the presence of $\boldsymbol{R}$, we only focused on the positive and negative labels, and missing labels did not lead to any loss.

In addition, a regularization term using a similarity matrix proposed by Yong et al. [7] was introduced to bring the latent feature vectors of drugs and targets with high similarity closer together. The drug neighborhood information was represented using an adjacency matrix $\mathbf{A}$, and its element was defined as follows:

$$\mathbf{A}_{i,\mu} = \begin{cases} S_{i,\mu}^d & \text{if } d_\mu \in N(d_i) \\ 0 & \text{otherwise,} \end{cases} \tag{2}$$

where $N(d_i)$ is constructed by selecting $K_1$ most similar drugs with $d_i$. Drug–target effect neighborhood information $\mathbf{B}$ could be defined similarly as follows:

$$\mathbf{B}_{j,v} = \begin{cases} S_{j,v}^t & \text{if } d_v \in N(t_j) \\ 0 & \text{otherwise.} \end{cases} \tag{3}$$

The objective function that minimized the distance between $d_i$ and its nearest neighbor $N(d_i)$ in the latent space was as follows:

$$\frac{\alpha}{2}\sum_{i=1}^{m}\sum_{\mu=1}^{m} \mathbf{A}_{i,\mu} \|\mathbf{u}_i - \mathbf{u}_\mu\|_F^2 = \frac{\alpha}{2}\text{tr}(\mathbf{U}^T \mathbf{L}^d \mathbf{U}), \tag{4}$$

where $\text{tr}(\cdot)$ is the trace of matrix, $\mathbf{u}_i$ is the i$^{\text{th}}$ row in $\mathbf{U}$. $\mathbf{L}^d = (\mathbf{D}^d + \breve{\mathbf{D}}^d) - (\mathbf{A} + \mathbf{A}^T)$, in which $\mathbf{D}^d$ and $\breve{\mathbf{D}}^d$ are defined as $\mathbf{D}_{ii}^d = \sum_{\mu=1}^{m} \mathbf{A}_{i,\mu}$ and $\breve{\mathbf{D}}_{\mu\mu}^d = \sum_{i=1}^{m} \mathbf{A}_{i,\mu}$, respectively. It was different from the ordinal graph Laplacian, which considered effects from all similar drugs and targets. The same could be formulated for the elements of the target action using $\mathbf{B}$ as follows:

$$\frac{\beta}{2}\sum_{i=1}^{m}\sum_{\mu=1}^{m} \mathbf{B}_{j,v} \|\mathbf{v}_j - \mathbf{v}_v\|_F^2 = \frac{\beta}{2}\text{tr}(\mathbf{V}^T \mathbf{L}^t \mathbf{V}). \tag{5}$$

By combining equations (1), (4), and (5), we proposed the NRBdMF algorithm. Its objective function was as follows:

$$\arg\min_{\mathbf{U},\mathbf{V}} J = \frac{1}{2}\|\boldsymbol{R} \circ (\boldsymbol{Y} - \boldsymbol{U}\boldsymbol{V}^T)\|_F^2 + \frac{1}{2}tr[\boldsymbol{U}^T(\lambda_d \boldsymbol{I} + \alpha \boldsymbol{L}_d)\boldsymbol{U}] + \frac{1}{2}tr[\boldsymbol{V}^T(\lambda_t \boldsymbol{I} + \beta \boldsymbol{L}_t)\boldsymbol{V}], \tag{6}$$

where $\boldsymbol{I}$ denotes the identity matrix, $\boldsymbol{L}_d$ and $\boldsymbol{L}_t$ are a kind of graph Laplacian that considers only neighborhood influence. In addition, $\lambda_d, \lambda_t, \alpha$, and $\beta$ are hyperparameters.

To estimate the latent feature matrices $\boldsymbol{U}$ and $\boldsymbol{V}$, we proposed an alternating gradient descent procedure. The partial gradients for $\boldsymbol{U}$ and $\boldsymbol{V}$ were as follows:

$$\frac{\partial J}{\partial \boldsymbol{U}} = \boldsymbol{R} \circ (\boldsymbol{U}\boldsymbol{V}^T - \boldsymbol{Y})\boldsymbol{V} + (\lambda_d \boldsymbol{I} + \alpha \boldsymbol{L}_d)\boldsymbol{U} \tag{7}$$

$$\frac{\partial J}{\partial \boldsymbol{V}} = \boldsymbol{R}^T \circ (\boldsymbol{V}\boldsymbol{U}^T - \boldsymbol{Y}^T)\boldsymbol{U} + (\lambda_t \boldsymbol{I} + \beta \boldsymbol{L}_t)\boldsymbol{V}. \tag{8}$$

Note that it was possible to assign different weights to each of the positive and negative labels with $\boldsymbol{R}$, allowing for bidirectional learning. We used the AdaGrad algorithm [21] to accelerate the convergence of the gradient descent optimization as shown in Algorithm 1.

**Algorithm 1 : NRBdMF Optimization Algorithm.**
**Input:** $\boldsymbol{Y}, \boldsymbol{S}^d, \boldsymbol{S}^t, pw, nw, K_1, K_2, \lambda_d, \lambda_t, \alpha, \beta, \gamma$;
**Output:** $\boldsymbol{U}, \boldsymbol{V}$.
1: Initialize $\boldsymbol{U}$ and $\boldsymbol{V}$ randomly, and set $\varphi_{ik} = 0, \phi_{jk} = 0$, $\forall 1 \leq i \leq m, 1 \leq j \leq n,$ and $1 \leq k \leq r$;
2: Construct $\boldsymbol{A}$ and $\boldsymbol{B}$ according to eqs. (2) and (3);
3: Compute $\boldsymbol{L}^d$ and $\boldsymbol{L}^t$ according to eqs. (4) and (5);
4: **for** $t = 1, \cdots, max\_iter$ **do**
5: $\quad \boldsymbol{G}^d \leftarrow \frac{\partial \boldsymbol{J}}{\partial \boldsymbol{U}}$;
6: $\quad$ **for** $i = 1, \cdots, m$ **do**
7: $\quad\quad$ **for** $k = 1, \cdots, r$ **do**
8: $\quad\quad\quad \varphi_{ik} \leftarrow \varphi_{ik} + g^d_{ik} \cdot g^d_{ik}$
9: $\quad\quad\quad u_{ik} \leftarrow u_{ik} - \gamma \frac{g^d_{ik}}{\sqrt{\varphi_{ik}}}$;
10: $\quad\quad$ **end for**
11: $\quad \boldsymbol{G}^t \leftarrow \frac{\partial \boldsymbol{J}}{\partial \boldsymbol{V}}$;
12: $\quad$ **for** $j = 1, \cdots, n$ **do**
13: $\quad\quad$ **for** $k = 1, \cdots, r$ **do**
14: $\quad\quad\quad \phi_{jk} \leftarrow \phi_{jk} + g^t_{jk} \cdot g^t_{jk}$
15: $\quad\quad\quad v_{jk} \leftarrow v_{jk} - \gamma \frac{g^t_{jk}}{\sqrt{\phi_{jk}}}$;
16: $\quad\quad$ **end for**
17: $\quad$ **end for**
18: $\quad$ **end for**
19: **end for**
20: **return** $\boldsymbol{U}, \boldsymbol{V}$

*2.6. Predicting the side effects of drugs using multilabel data*

*2.6.1. Drug side effect matrix*

The SIDER database (version 4.1) [22] was used to define the presence of drug side effects. We extracted side effects that were registered in the MedDRA preferred terms (MedDRA Version 21.1)[23]. In total, 141,809 associations between 1429 drugs and 4138 side effects were obtained.

*2.6.2. Drug indication matrix*

The SIDER database also contains information regarding drug indications. As with drug side effects, we focused on the MedDRA preferred terms and extracted drug indication pairs.

The obtained dataset contained 14,535 drug indication associations between 1423 drugs and 2154 indications.

*2.6.3. Multilabel matrix*

We extracted the drugs and diseases that were present in the intersection of the above two matrices. After assigning labels of +1 and −1 to the side effects and indications, respectively, the two matrices were added together to define a multilabel matrix that reflected the bidirectional nature of drug effects. This matrix contained +1, 0, and −1 labels. The labels +1 and −1 indicated the side effects and indications, respectively, while the label 0 indicated the missing value or the contradictory relationship where both side effects (+1 label) and indications (−1 label) were registered in SIDER and added up to 0. Finally, we defined a relational matrix between 646 drugs and 499 diseases, including 17,546 positive and 2343 negative labels.

*2.7. Performance evaluation of NRBdMF*

We demonstrated the effectiveness of preparing multilabel data to predict drug side effects by comparing our proposed NRBdMF with NRLMF using 10-fold CV. Data splitting between training datasets (80%), validation datasets (10%), and test datasets (10%) was performed using random selection under the CVS2 and CVS3 conditions. Although a difference existed between NRBdMF and NRLMF in that the data used for input were multilabel and binary, the relationships used for training, validation, and testing were the same because the seed value was fixed during division.

Both methods introduced a neighborhood regularization, and the same values were used for the associated hyperparameters, $\lambda_d, \lambda_t, \alpha, \beta, K_1,$ and $K_2$. Specifically, we set $\lambda_d = \lambda_t = 0.65$, $\alpha = \beta = 0.1$, and $K_1 = K_2 = 5$ referring to the initial values of NRLMF. Label weights were assigned using a grid search for each method because NRBdMF allows different weights to be assigned for the positive and negative directions. For NRBdMF, both *pw* and *nw*, which were the weights of the positive and negative labels, were selected from $\{10^{-1}, 10^0, 10^1\}$. For NRLMF, c was a parameter regarding weight of the observed interaction and was selected from $\{10^0, 10^1, 10^2\}$. For both NRBdMF and NRLMF, tolerance was selected from $\{10^{-5}, 10^{-3}, 10^{-1}\}$.

We defined the enrichment score (ES) for predicting side effects using the respective model optimized by the grid search. First, we calculated enrichment values (EV) for each of the side effects and indications with reference to the gene set variation analysis score [24] as follows:

$$EV = |KS^+| - |KS^-| \qquad (9)$$

where $KS^+$ and $KS^-$ are Kolmogorov–Smirnov (KS) scores, indicating the largest positive and negative deviations from zero, respectively. A higher value indicates that the targets are enriched in the top-ranked predictions. In the present study, the difference of EV scores between side effects and indications was defined as ES:

$$ES = EV^{side\ effects} - EV^{indications}. \qquad (10)$$

## 3. Results

*3.1. Literature survey*

In January 2022, we conducted a literature survey on matrix factorization algorithms to predict the relationship between drugs and their effects. Papers published before 2009 were excluded from the study selection because the matrix factorization method was first devised as a recommendation system in 2009 [25]. After eliminating studies based on titles, 79 studies were left. Forty-one of these studies were excluded after reading the full paper and 38 were finally selected for review.

By surveying the algorithms used in the selected 38 studies, the existing algorithms were broadly classified into three categories based on the number of matrices to be decomposed (Fig. 1). Among them, we found seven representative algorithms (NMF, GRNMF, LMF, NRLMF, CMF, TMF, and IMC) [4–10], which have been used widely since their development.

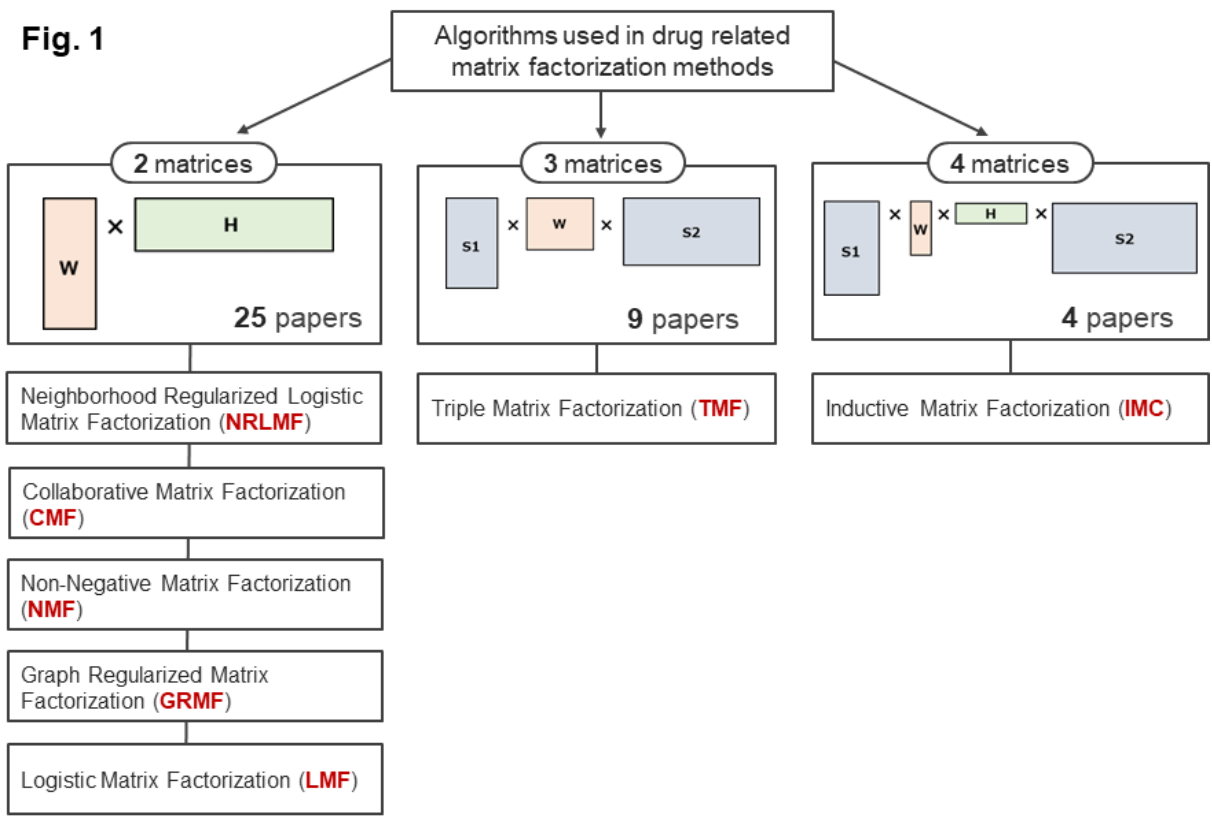

**Fig. 1.** Classification of the algorithms used in 38 studies on drug-related matrix factorization methods. We selected seven representative algorithms.

*3.2. Performance evaluation of representative algorithms*

Next, we conducted performance evaluation of the seven representative algorithms using CV methods (CVS1, CVS2, and CVS3) that are widely employed in the field of recommendation systems (Fig. 2a) [2,3,16,26]. NMF, GRNMF, and LMF require no prior information, called as a kernel, other than the matrix to be trained. NRLMF and CMF need the input of similarity square matrix information as kernels for each row and column. In addition, TMF and IMC require the low-dimensional vector that is obtained from dimensionality reduction. All datasets and kernels used for comparison in the present study will be made available for public access at https://github.com/mizuno-group/NRBdMF.

The performance of the seven representative algorithms was evaluated using two types of benchmark datasets by 10-fold CV performed five times in the three scenarios. The area under precision-recall curve (AUPR) of each method in predicting drug–target proteins is shown in Fig. 2b. NRLMF consistently showed the best results in all three scenarios. Regarding another benchmark dataset, the drug–disease dataset, NRLMF was the best in CVS1, while TMF outperformed the other methods in CVS2 and CVS3 (Fig. 2c). These findings suggest that NRLMF was robust and superior among the seven representative existing algorithms compared in the present study. Similarly, the prediction performance was evaluated in terms of the area under receiver operating characteristic curve (AUROC; Supplementary Fig. S1). In addition, the PR and ROC curves for the first CV (out of five times) are shown in Supplementary Figs S2 and S3. The overall summary of the AUPR and AUROC results for each benchmark dataset is shown in Table 2.

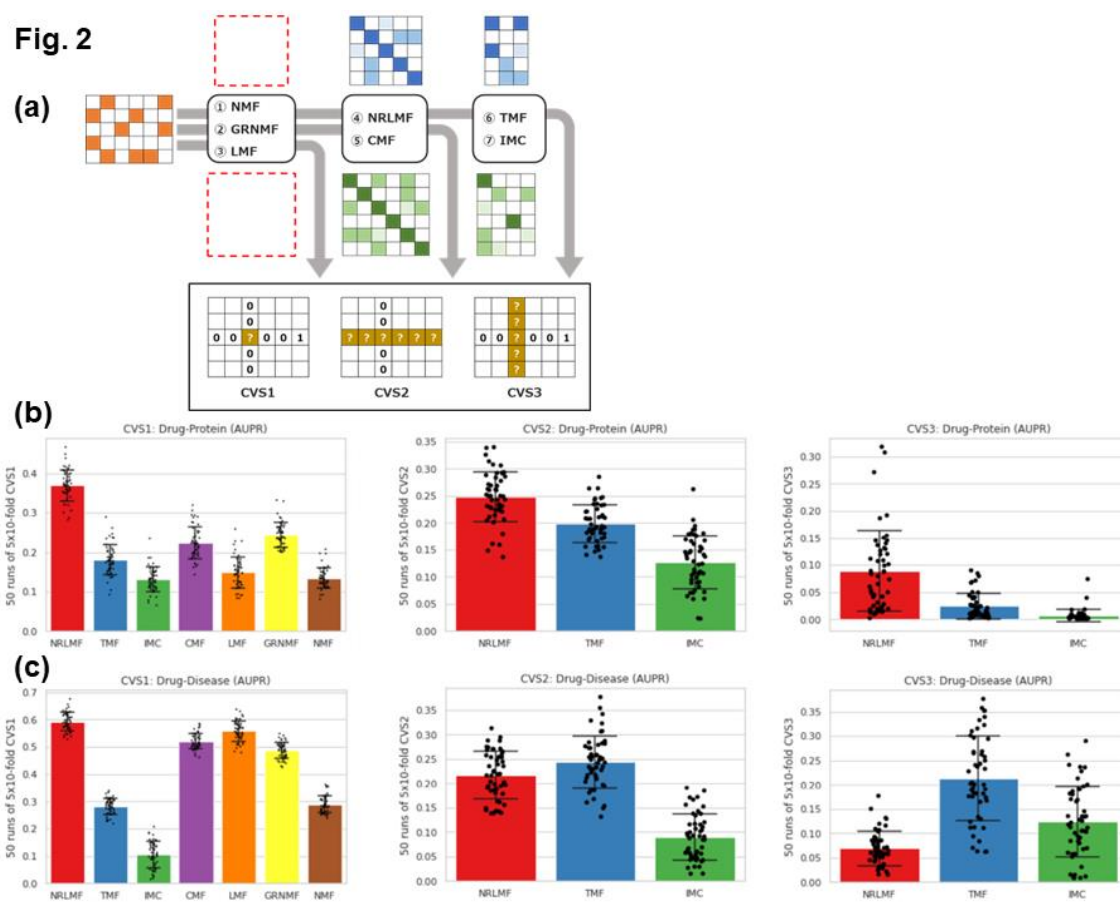

**Fig. 2.** Performance comparison of representative algorithms. (a) Overview of the evaluation system for the seven representative algorithms. (b) area under precision-recall curve (AUPR) values in the 50 runs of 10-fold cross-validation (CV) performed five times under three different settings (CVS1, CVS2, and CVS3) for the drug–protein benchmark dataset. (c) AUPR values in the 50 runs of 10-fold CV performed five times under the settings CVS1, CVS2, and CVS3 for the drug–disease benchmark dataset. (c) AUPR values in the 50 runs of 10-fold CV performed five times under the settings CVS1, CVS2, and CVS3 for the drug–disease benchmark dataset.

**Table 2.** AUPR and AUROC values of seven representative algorithms.

|  |  | NRLMF | TMF | IMC | CMF | LMF | GRNMF | NMF |
|---|---|---|---|---|---|---|---|---|
|  |  | colspan Drug-Protein | | | | | | |
| AUPR | CVS1 | **0.370±0.039** | 0.181±0.039 | 0.131±0.031 | 0.223±0.040 | 0.149±0.039 | 0.244±0.031 | 0.131±0.026 |
|  | CVS2 | **0.248±0.046** | 0.198±0.034 | 0.127±0.049 | - | - | - | - |
|  | CVS3 | **0.0892±0.074** | 0.0245±0.023 | 0.00628±0.011 | - | - | - | - |
| AUROC | CVS1 | **0.914±0.017** | 0.892±0.017 | 0.814±0.036 | 0.77±0.030 | 0.815±0.027 | 0.707±0.027 | 0.787±0.020 |
|  | CVS2 | 0.807±0.043 | **0.872±0.022** | 0.808±0.033 | - | - | - | - |
|  | CVS3 | 0.7478±0.086 | **0.7491±0.054** | 0.542±0.078 | - | - | - | - |
|  |  | Drug-Disease | | | | | | |
| AUPR | CVS1 | **0.592±0.033** | 0.282±0.030 | 0.104±0.049 | 0.521±0.029 | 0.557±0.038 | 0.486±0.030 | 0.289±0.031 |
|  | CVS2 | 0.217±0.048 | **0.243±0.052** | 0.0876±0.046 | - | - | - | - |
|  | CVS3 | 0.0632±0.035 | **0.213±0.086** | 0.124±0.072 | - | - | - | - |
| AUROC | CVS1 | 0.889±0.015 | **0.897±0.013** | 0.695±0.048 | 0.881±0.014 | 0.865±0.018 | 0.838±0.016 | 0.856±0.015 |
|  | CVS2 | 0.813±0.037 | **0.865±0.026** | 0.666±0.058 | - | - | - | - |
|  | CVS3 | 0.644±0.058 | **0.838±0.039** | 0.690±0.063 | - | - | - | - |

*3.3. NRBdMF for predicting drug side effects*

Drug effects in individuals have two opposite effects: namely, therapeutic effects and side effects. In this regard, the binary representation used in the field of recommendation systems is not suitable for describing drug effects and directionality should be considered in the application of recommendation systems for the prediction of drug effects, which would make the model richer in representation and more interpretable [12]. However, most existing recommendation algorithms take only binary information as the target matrix [2,3,16,26] and, to the best of our knowledge, no studies that predict drug side effects using a recommendation system that considers the bilateral character of drug effects (i.e., a therapeutic effect one time and a side effect the other time) are available.

In the present study, we addressed the abovementioned issues as follows. First, we prepared a multilabel dataset in which existing side effect information was assigned the +1 label, existing indication information was assigned the −1 label, and unknown or contradictory relationships were assigned the 0 label. Using this multilabel matrix, we could predict drug side effects while considering the bidirectional nature of drug effects, such as side effects and indications. Then, we developed a more generalized algorithm that could be applied to the multilabel data, inspired by NRLMF, which was one of the best performers (Fig. 2). The overview of the side effect prediction using NRBdMF is shown in Fig. 3.

**Fig. 3**

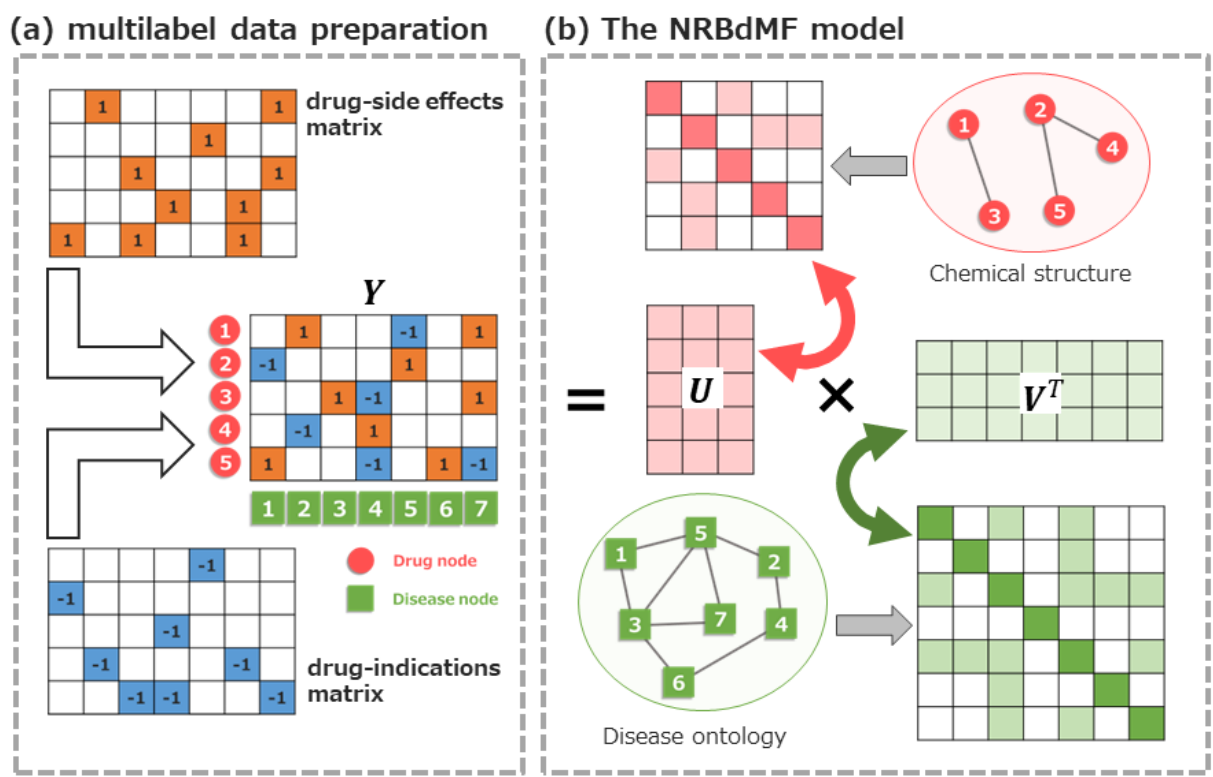

**Fig. 3.** The overall workflow for predicting side effects using neighborhood regularized bidirectional matrix factorization (NRBdMF). (a) Preparation of multilabel input dataset with merged information on side effects and indications. 1 and −1 indicate the presence of side effects and indications, respectively. (b) The model of neighborhood regularized logistic matrix factorization (NRLMF). We prepared similarity square matrices for drugs and diseases, respectively. Then, neighborhood regularization using the local structure of the interaction was performed.

*3.4. Comparison with NRLMF*

We compared the performance of our proposed multilabel learning method with that of the conventional binary learning method for predicting side effects. Assuming the more realistic task of predicting new drugs and unknown side effects, we used the CVS2 and CVS3 settings.

We split data between the training (80%), validation (10%), and test datasets (10%) using random selection for 10-fold CV. The parameters of each model were tuned via a grid search using the training and validation datasets, and optimized models were prepared. Parameter robustness using the grid search is presented in Supplementary Fig. S4. The prediction performance was evaluated for the remaining test datasets using the optimized version of each model.

In predicting the presence or absence of drug side effects, we evaluated the enrichment of known side effects (+1 label) and indications (−1 label). In an ideal drug side effect prediction, the positive labels are enriched, and the negative labels are sparse in the top-ranked predictions. A representative result of enrichment of the two conflicting labels (side effects and indications) is shown in Fig. 4a. The positive labels for side effects, indicated by orange lines, were highly enriched in both NRLMF and NRBdMF. The indication labels, indicated by blue lines, were concentrated at the top as well as side effects in NRLMF while NRBdMF succeeded in pushing them to the bottom.

The difference between the enrichment of side effects and the enrichment of indications is shown as an ES in Fig. 4b. In CVS3, the mean ES of NRBdMF (0.588 ± 0.101) outperformed that of NRLMF (0.191 ± 0.106). This indicates that the proposed method, NRBdMF, can enrich side effects and eliminate indications in the top-ranked predictions, which reduces the false positives and provides interpretable prediction results. The AUPR and AUROC of each NRLMF and NRBdMF are shown in Supplementary Fig. S5 and only limited difference was observed between the two methods. These findings were consistent for CVS2, with the NRBdMF score (0.833 ± 0.0949) outperforming the NRLMF score (0.376 ± 0.0526; Supplementary Fig. S6).

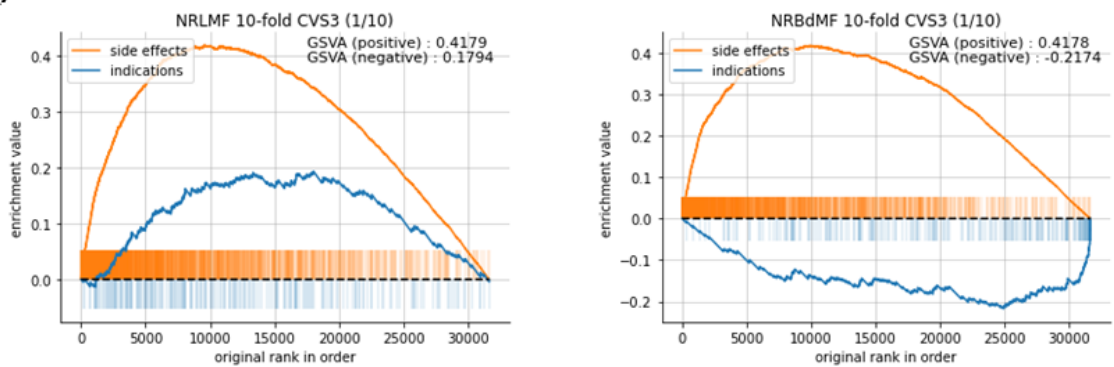

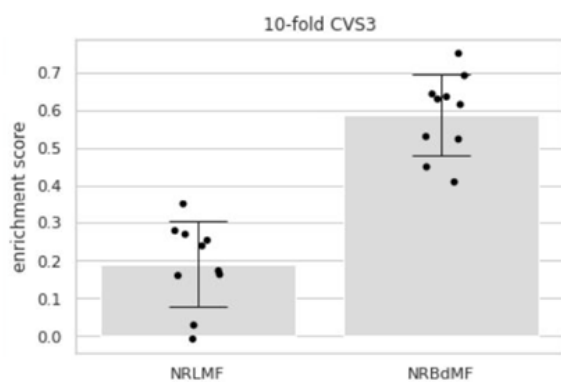

**Fig. 4.** Comparison of side effect prediction performance between NRBdMF and NRLMF. (a) Enrichment of known side effects (orange) and indications (blue) when predicting drug side effects using NRLMF and NRBdMF under CVS3. (b) Enrichment score for each model. Enrichment score was defined using the difference in enrichment values for each of the side effects and indications. Enrichment values were calculated with reference to the statistics proposed in the gene set variation analysis (GSVA). The mean enrichment scores were 0.191 ± 0.106 and 0.588 ± 0.101 for NRLMF and NRBdMF, respectively. ±, standard deviation.

*3.5. Case studies*

NRBdMF eliminated more false positives as a whole under the CVS2 and CVS3 settings; therefore, we examined case studies individually in more detail. In the case studies, we specifically compared the output of NRLMF and NRBdMF side effect predictions targeting hypertension, which is a serious side effect. We trained both prediction models using all known interactions except for the targe disease in the multilabel matrix, and then predicted drug candidates that would cause the disease as a side effect.

The top 10 and bottom 10 drug candidates that could cause hypertension predicted using NRBdMF and NRLMF are shown in Table 3. The NRBdMF output showed enrichment of side effects (positive labels) at the top and indications (negative labels) at the bottom. NRLMF output clearly contained false positives, with drugs known to treat hypertension, such as bosentan and nifedipine, being the top candidates detected as producing side effects [27,28]. Regarding the bottom-ranked predictions, the NRBdMF output included statins, which are drugs that treat hypertension, while the NRLMF output included several drugs with known side effects. This finding indicates that our proposed NRBdMF model can enrich more plausible side effect candidates at the top ranks and less plausible candidates at the bottom ranks in predicting drug side effects. This finding was true not only for case studies focusing on disease, but also for specific drugs such as doxorubicin (Table S3)[29,30].

**Table 3.** Top 10 and bottom 10 predicted drugs that produce hypertension as side effects. +, known side effects; −, known indications; N.A., evidence is not available in SIDER.

| Rank | NRBdMF Score | Name | Label in SIDER | NRLMF Score | Name | Label in SIDER |
|---|---|---|---|---|---|---|
| 1 | 0.128088 | paroxetine | + − | 0.282737 | paroxetine | + − |
| 2 | 0.124531 | pregabalin | + | 0.202664 | cyclophosphamide | + |
| 3 | 0.121634 | ropinirole | + − | 0.162674 | fluoxetine | + |
| 4 | 0.117924 | celecoxib | + − | 0.138129 | citalopram | + |
| 5 | 0.116124 | pramipexole | + | 0.104735 | pregabalin | + |
| 6 | 0.113656 | valdecoxib | + | 0.081821 | tacrolimus | + − |
| 7 | 0.11164 | methotrexate | N.A. | 0.057096 | ifosfamide | + |
| 8 | 0.106965 | aripiprazole | + | 0.044413 | sibutramine | + |
| 9 | 0.103946 | naproxen | + | 0.036354 | bosentan | − |
| 10 | 0.103685 | tramadol | + | 0.035278 | nifedipine | − |
| | | | | | | |
| 637 | -0.09728 | warfarin | N.A. | 7.73E-05 | fluphenazine | + |
| 638 | -0.10752 | prazosin | − | 7.32E-05 | stavudine | N.A. |
| 639 | -0.1077 | nitroglycerin | − | 7.00E-05 | mitotane | + |
| 640 | -0.11218 | famotidine | − | 6.56E-05 | vincristine | + |
| 641 | -0.11726 | rosuvastatin | N.A. | 6.34E-05 | modafinil | + |
| 642 | -0.11903 | spironolactone | − | 5.87E-05 | mitoxantrone | + |
| 643 | -0.13723 | fluvastatin | + − | 5.09E-05 | travoprost | + |
| 644 | -0.1453 | pravastatin | + − | 4.86E-05 | rifaximin | **N.A.** |
| 645 | -0.14678 | simvastatin | − | 4.85E-05 | clindamycin | + |
| 646 | -0.15564 | lovastatin | − | 3.04E-05 | tinidazole | **N.A.** |

## 4. Discussion

In the present study, we surveyed drug effect prediction using matrix factorization and compared the performance of existing representative algorithms. Furthermore, we proposed NRBdMF, which is a more general algorithm based on the existing superior method (NRLMF)[7]. We achieved more interpretable side effect prediction by applying this proposed method to a multilabel dataset considering the two aspects of drug side effects and therapeutic effects as directionality.

The performance of representative algorithms was evaluated using AUPR. However, not as much difference was observed between the methods when AUROC was used as that observed with AUPR (Supplementary Fig. S1). This phenomenon with AUROC is known to occur when the number of positive labels is much lesser than that of negative labels, and AUPR is more appropriate than AUROC [15,31,32]. Through these comparisons, NRLMF was found to perform better in CVS1 than in CVS2 and CVS3, which was consistent with the findings of the original study reviewed in the present study [7].

Throughout the survey, most algorithms used only the presence or absence of known relationships as binary information. This is also true for NRLMF, which showed excellent performance and used logistic regression to assume that the labels fit between 0 and 1. While most of these methods cannot be applied directly to the multilabel data prepared in the present study, our proposed NRBdMF method is more flexible to the type of the input data.

In fact, the proposed method was successful in predicting drug side effects by considering the bidirectional nature of drug effects and concurrently enriching the relationship of known side effects and that of indications at the top and bottom of the prediction list, respectively. The case studies showed that this method can avoid false positives in both the prediction of side effects of new drugs (CVS2) and drugs causing unknown side effects (CVS3).

The present study has some potential limitations. We obtained data on the relationships between 1423 drugs and 2154 adverse drug reactions and between 1429 drugs and 4138 indications from SIDER [22]. However, the kernels of drugs and diseases that needed to be input into the matrix factorization process were scarce, and the final analysis included 646 drugs and 499 diseases relationships, which limited the scope of target prediction. In future, the scope of what the kernel reflects must be expanded. In addition, the definition of relationships that relies on a single database, SIDER [22], may be insufficient in terms of the amount of information. Furthermore, the collection and use of a wider range of side effect and indication relationships by aggregating existing findings and the contents of package inserts through natural language processing and knowledge graphs are warranted [33–35].

In the present study, the positive and negative effects of drugs were considered bidirectional, and we prepared symmetrical data with respect to the center as 0. Essentially, this bidirectional multilabel data can be extended to ordinal scale data and our proposed NRBdMF method can also be applied to these data. Even if we input multiclass information, such as clinical scores or frequencies, as ordinal scale data [36,37], our proposed method is expected to work well as a tool to obtain novel findings.

The concept of predicting adverse drug reactions by considering the duality of drug efficacy as directionality can also be applied to other algorithms. TMF performed next to NRLMF in this validation; therefore, a more generalized algorithm for TMF that can handle multiple labels and analyze it in the same way should be developed in the future. Moreover, many methods using neural network algorithms have been devised in this field [38–41]. Although we did not summarize them in the present study because we focused on the conceptual aspects of matrix factorization, our directionality concept in predicting drug effects would be applicable to matrix factorization methods using neural network algorithms, which would be interesting to study in the future.

**STATEMENT OF SIGNIFICANCE**

**Problem or Issue**

Prediction of drug effects using a recommendation system based on matrix factorization has been well studied in recent years. Many methods have been devised, but the conceptual differences are not organized in concepts and do not consider directionality of drug effects, including therapeutic effects and adverse effects.

**What is Already Known**

Comparisons of the performance of existing recommendation systems are performed whenever a new algorithm is developed, and the superiority of some particular algorithm, such as NRLMF, is mentioned in some studies.

**What this Paper Adds**

We proposed a novel algorithm named NRBdMF, which can handle categorical data showing positive and negative bidirectionality, such as labels for drug effects (e.g., therapeutic and side effects). In predicting drug side effects, NRBdMF reduced false positives derived from the bilateral character of drug effects and improved the interpretability of results.

**CRediT authorship contribution statement**

**Iori Azuma:** Methodology, Software, Investigation, Writing – Original Draft, Visualization. **Tadahaya Mizuno:** Conceptualization, Resources, Supervision, Project administration, Writing – Review & Editing, Funding acquisition. **Hiroyuki Kusuhara:** Writing – Review & Editing

**Acknowledgments**

We thank all those who contributed to the construction of the following datasets employed in the present study: DTINet [18], MBiRW [19], LRSSL [20], and SIDER [22].

**Figure S1**

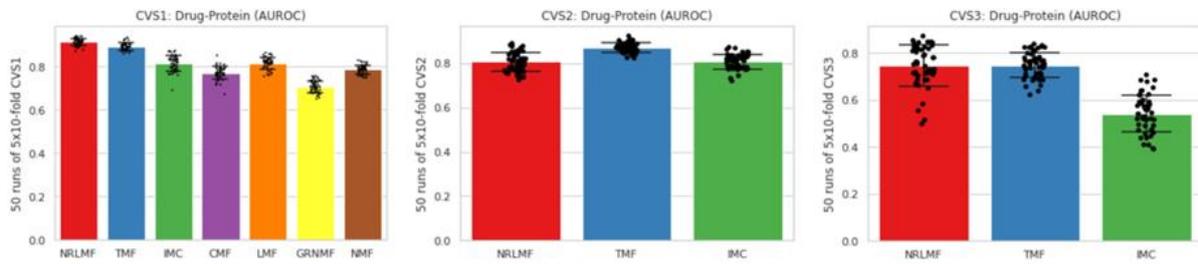

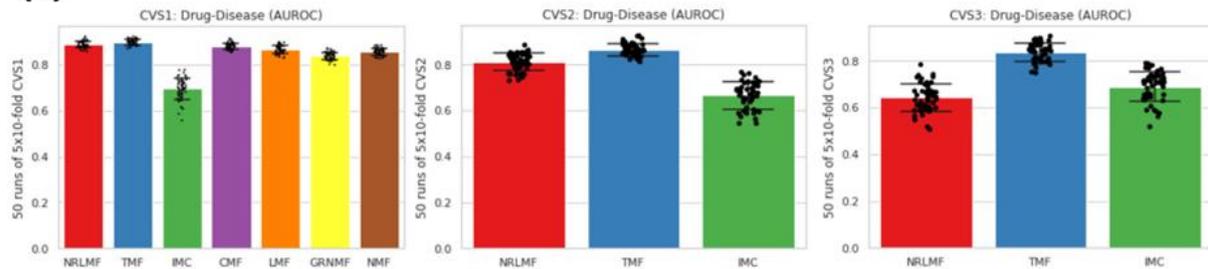

**Figure S1.** AUROC values in the 50 runs of 5 times 10-fold CVS1, CVS2 and CVS3. (a) Drug-Protein benchmark dataset. (b) Drug-Disease benchmark dataset.

**Figure S2**

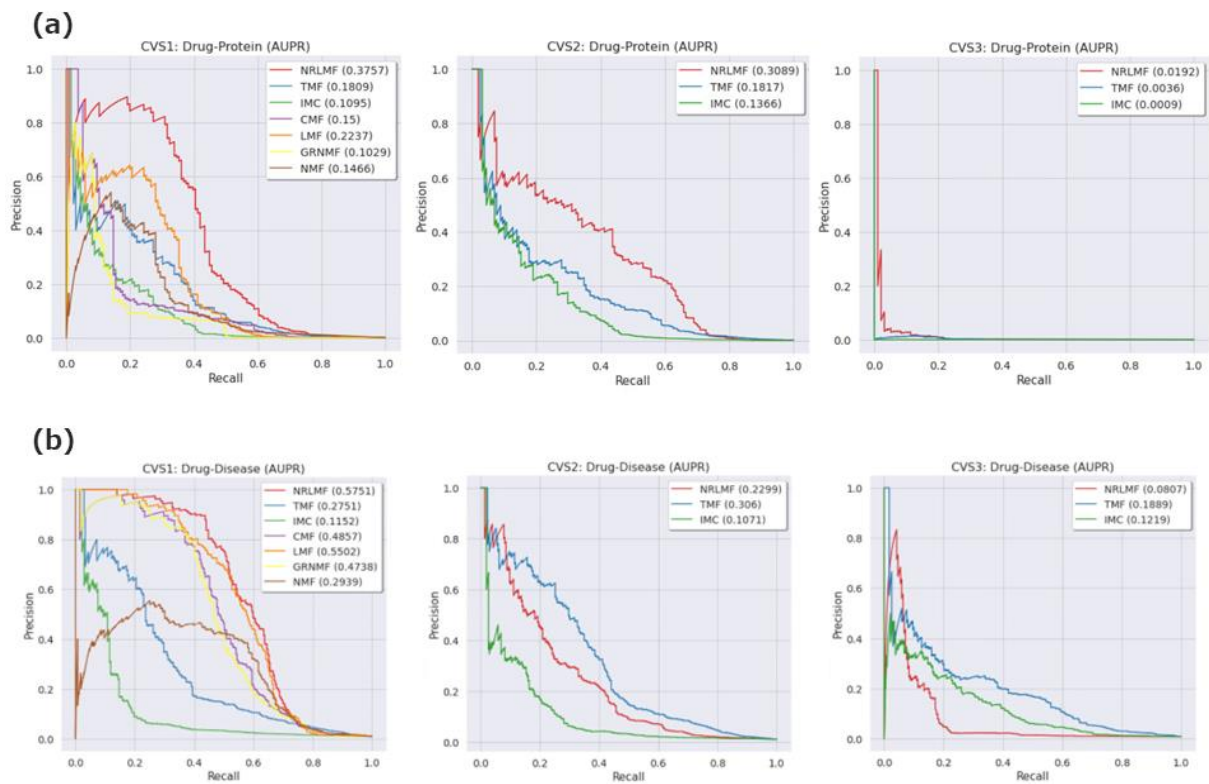

**Figure S2.** PR curves for the first 10-fold cross-validation out of 5 times. (a) Drug-Protein benchmark dataset. (b) Drug-Disease benchmark dataset.

**Figure S3**

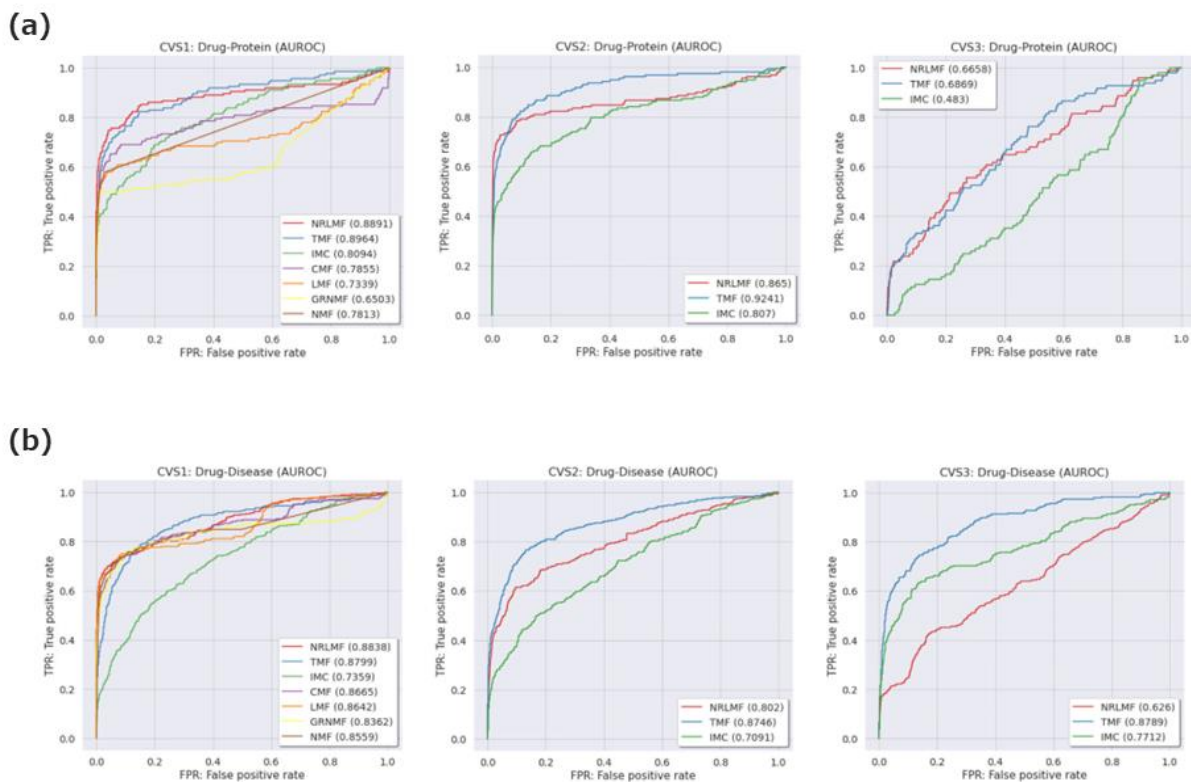

**Figure S3.** ROC curves for the first 10-fold cross-validation out of 5 times. (a) Drug-Protein benchmark dataset. (b) Drug-Disease benchmark dataset.

**Figure S4**

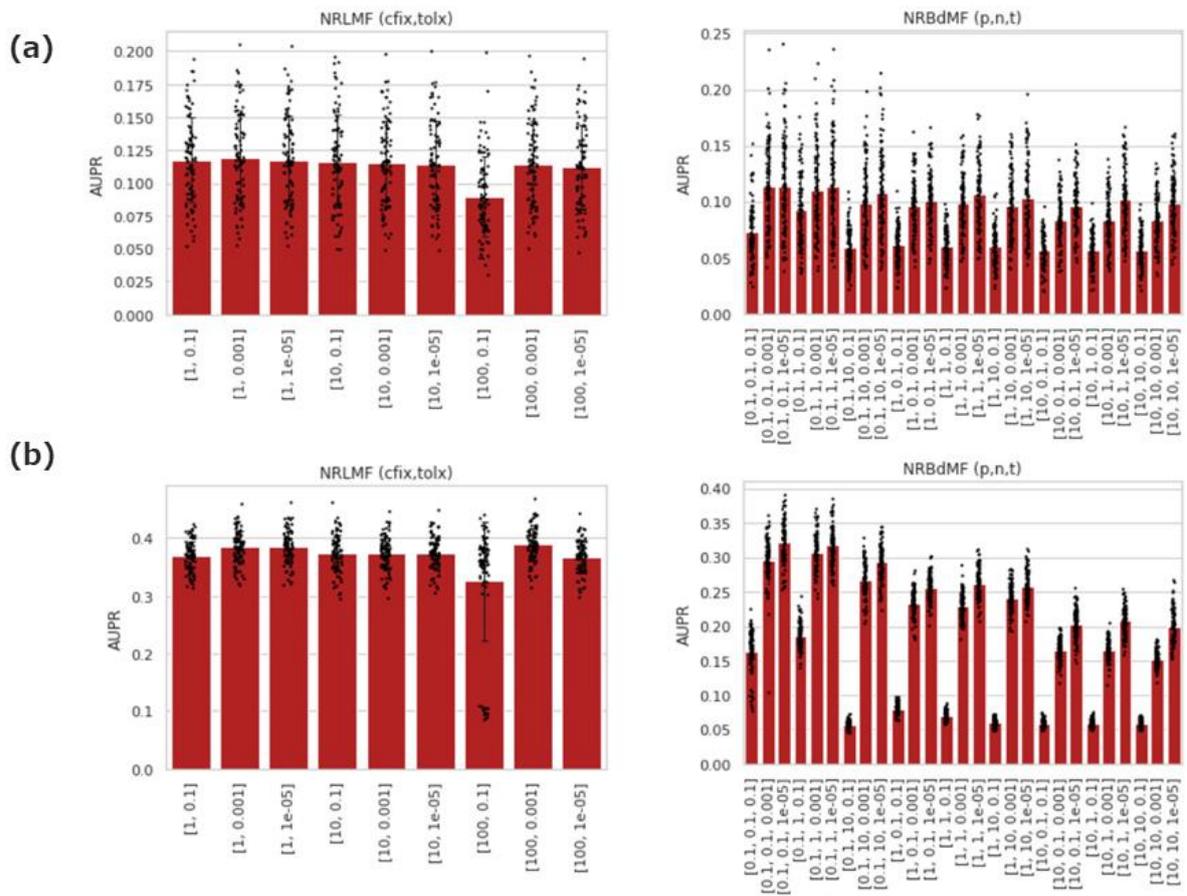

**Figure S4.** Tuning of hyperparameter by grid search. The tuned model was used to evaluate the prediction performance on the test dataset. (a) Under CVS2. (b) Under CVS3.

**Figure S5**

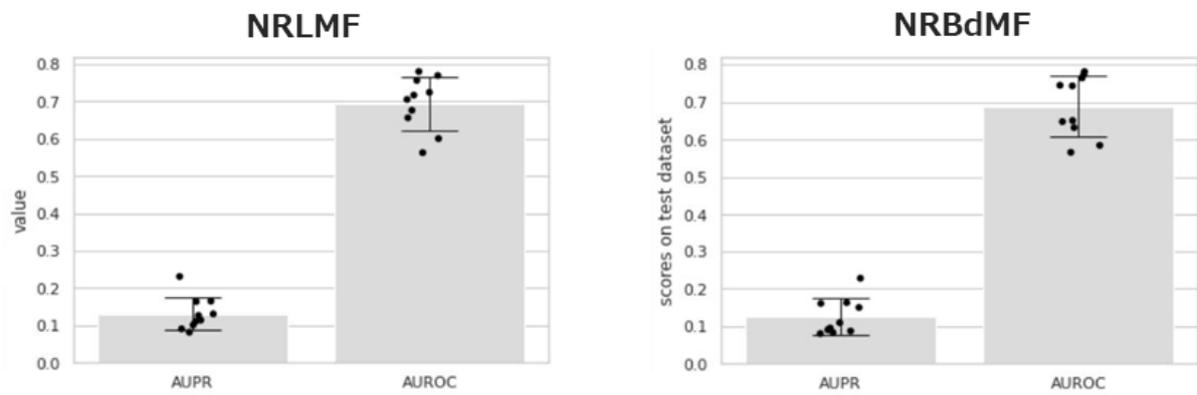

**Figure S5.** AUPR and AUROC for predicting side effects using NRLMF and NRBdMF under CVS3. On the left are the results of the NRLMF, with mean AUPR and AUROC values of 0.126±0.0390 and 0.701±0.0617, respectively. On the right are the results of the NRBdMF and the mean AUPR and AUROC values are 0.124±0.0463 and 0.688±0.0772, respectively. ±, standard deviation.

**Figure S6**

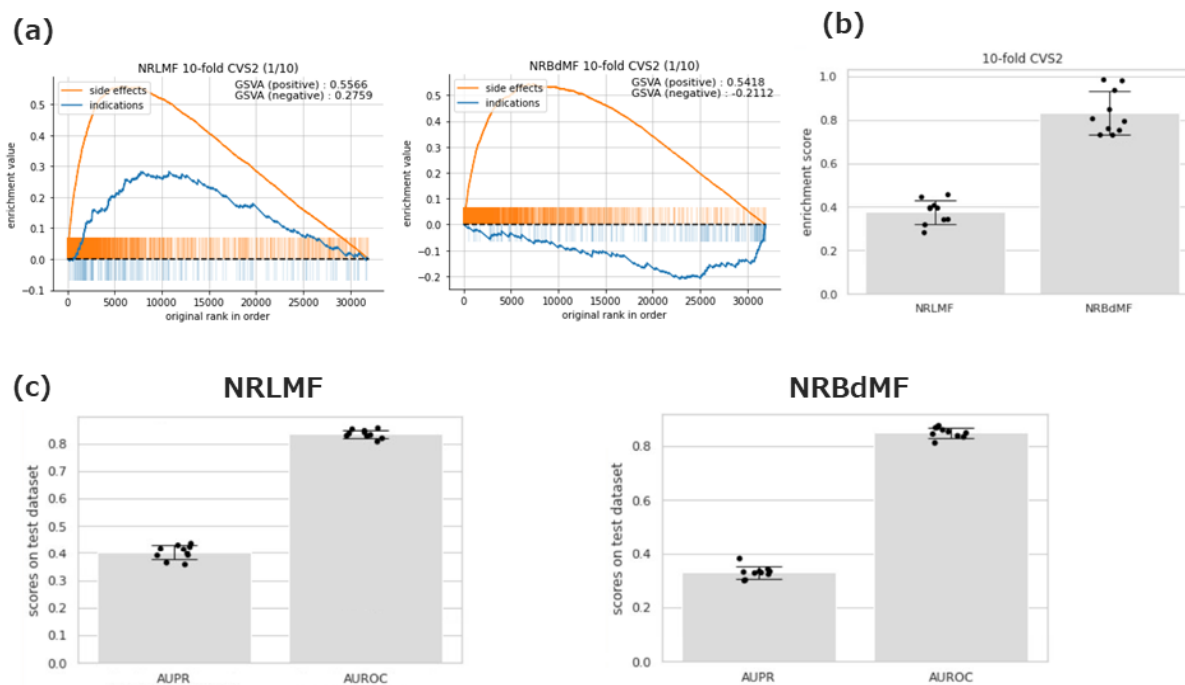

**Figure S6.** Comparison of side effect prediction performance between NRBdMF and NRLMF under CVS2. (a) Enrichment of known side effects (orange) and indications (blue) when predicting drug side effects with NRLMF and NRBdMF under CVS2. (b) Enrichment score for both models. Enrichment score was defined by the difference in enrichment values for each of the side effects and indications. The mean values are 0.376 ± 0.0526 and 0.833 ± 0.0949 for NRLMF and NRBdMF, respectively. (c) AUPR and AUROC for predicting side effects using NRLMF and NRBdMF under CVS2. On the left are the results of the NRLMF, with mean AUPR and AUROC values of 0.402 ± 0.0254 and 0.834 ± 0.0143, respectively. On the right are the results of the NRBdMF and the mean AUPR and AUROC values are 0.330 ± 0.0214 and 0.850 ± 0.0182, respectively. ±, standard deviation.

**Table S1**

**Table S1.** Research paper selection criterion.

| Steps | Parameters | Selection criteria |
|---|---|---|
| 1 | Keyword | Research papers contains keywords such as "drug", "interact", "predict", "matrix", "factorization", "completion" and "decomposition". |
| 2 | Duration | Research papers published after 2009. |
| 3 | Title | Research involves in predicting drug effects. |
| 4 | Full paper reading | Research papers developing a novel algorithm for predicting drug effects. |

**Table S2**

Table S2. Summary of the benchmark datasets.

|  | Drug-Protein | Drug-Disease |
|---|---|---|
| Number of drugs | 708 | 663 |
| Number of targets | 1512 | 409 |
| Number of interaction pairs | 1332 | 2532 |
| Sparsity of the interaction matrix | 0.124% | 0.934% |

## Table S3

Table S3. Top 10 and bottom 10 predicted diseases that will be produced by doxorubicin as side effects. +, known side effects; −, known indications; N.A., evidence is not available in SIDER.

| | NRBdMF | | | NRLMF | | |
|---|---|---|---|---|---|---|
| Rank | Score | Name | Label | Score | Name | Label |
| 1 | 0.990499 | Rash | + | 0.999992 | Dermatitis | + |
| 2 | 0.981601 | Constipation | + | 0.999989 | Rash | + |
| 3 | 0.977036 | Dermatitis | + | 0.99987 | Thrombocytopenia | + |
| 4 | 0.957277 | Alopecia | + | 0.999768 | Alopecia | + |
| 5 | 0.944795 | Angioedema | N.A. | 0.999376 | Leukopenia | + |
| 6 | 0.944043 | Thrombocytopenia | + | 0.999133 | Hypersensitivity | + − |
| 7 | 0.939247 | Leukopenia | + | 0.998993 | Anaemia | + |
| 8 | 0.917137 | Hypersensitivity | + − | 0.998957 | Constipation | + |
| 9 | 0.916272 | Agranulocytosis | + | 0.997135 | Myocardial infarction | N.A. |
| 10 | 0.909092 | Photosensitivity reaction | + | 0.995666 | Hypertension | + |
| 490 | -0.23734 | Non-small cell lung cancer | − | 8.70E-08 | Nephritis | N.A. |
| 491 | -0.24664 | Schizoaffective disorder | N.A. | 8.26E-08 | Iritis | N.A. |
| 492 | -0.30244 | Breast cancer | − | 6.09E-08 | Angle closure glaucoma | N.A. |
| 493 | -0.30839 | Immunodeficiency | + − | 5.48E-08 | Glaucoma | N.A. |
| 494 | -0.31512 | Cervix carcinoma | N.A. | 4.28E-08 | Lactic acidosis | + |
| 495 | -0.33811 | Hodgkin's disease | − | 4.15E-08 | Polymyalgia rheumatica | N.A. |
| 496 | -0.33994 | Type 2 diabetes mellitus | N.A. | 3.85E-08 | Hepatitis A | N.A. |
| 497 | -0.34626 | Hodgkin's disease lymphocyte | − | 2.53E-08 | Uveitis | N.A. |

|     |          | predominance type stage unspecified                                       |          |          |              |          |
| --- | -------- | ------------------------------------------------------------------------- | -------- | -------- | ------------ | -------- |
| 498 | -0.34627 | Hodgkin's disease lymphocyte depletion type stage unspecified             | －       | 1.94E-08 | Lichen planus | **N.A.** |
| 499 | -0.34884 | Schizophrenia                                                             | **N.A.** | 1.41E-08 | Fibromyalgia | **N.A.** |